# Evaluating the Performance of BSBL[1] Methodology for EEG Source Localization

## On a Realistic Head Model

Sajib Saha[1, 3], Rajib Rana[5], Ya.I. Nesterets[1,4], M. Tahtali[3], Frank de Hoog[2], and T.E. Gureyev[1,4]

1) CSIRO Materials Science and Engineering, Clayton, VIC 3168, Australia
2) CSIRO Computational Informatics, Canberra, ACT
3) University of New South Wales, Canberra, ACT 2610, Australia
4) University of New England, Armidale, NSW 2351, Australia
5) University of Southern Queensland, QLD 4350, Australia

**Abstract:**

Source localization in EEG represents a high dimensional inverse problem, which is severely ill-posed by nature. Fortunately, sparsity constraints have come into rescue as it helps solving the ill-posed problems when the signal is sparse. When the signal has a structure such as block structure, consideration of block sparsity produces better results. Knowing sparse Bayesian learning is an important member in the family of sparse recovery, and a superior choice when the projection matrix is highly coherent (which is typical the case for EEG), in this work we evaluate the performance of block sparse Bayesian learning (BSBL) method for EEG source localization. It is already accepted by the EEG community that a group of dipoles rather than a single dipole are activated during brain activities; thus, block structure is a reasonable choice for EEG. In this work we use two definitions of blocks: Brodmann areas and automated anatomical labelling (AAL), and analyze the reconstruction performance of BSBL methodology for them. A realistic head model is used for the experiment, which was obtained from segmentation of MRI images. When the number of simultaneously active blocks is 2, the BSBL produces overall localization accuracy of less than 5 mm without the presence of noise. The presence of more than 3 simultaneously active blocks and noise significantly affect the localization performance. Consideration of AAL based blocks results more accurate source localization in comparison to Brodmann area based blocks.

**Introduction:**

The problem of source localization in EEG has gained significant attention in recent years because of its potential diagnostic value for epilepsy [2], stroke [3, 4], traumatic brain injury [5] and other brain disorders. Localization of the sources of electrical activity inside the brain works by measuring the scalp potentials produced by the electric activity in the brain (this is modelled using electric current dipoles), and then working back and estimating the dipoles that best fit the measurements. The EEG source localization represents a high-dimensional inverse problem, which is severely ill-posed [6] by nature and has an infinite number of solutions [7]. In order to find an appropriate unique solution, among the set of possible ones, constraints are applied into the problem [8]. In the literature, the most commonly used constraint is the minimum-norm constraint [9-11], which finds the solution that best matches the measurements with the smallest $l_2$ residuals. The strength of $l_2$ norm based approaches is their low computational cost; however such methods are often criticized for generating very broadly distributed or "smeared" sources in the reconstruction region [12] and for poor performance for multiple simultaneously active sources [13, 14].

In the last two decades, significant efforts have been made to solve ill-posed problems using sparse priors. Techniques relying on sparseness constraint search for a solution vector that not only matches the measurements, but also has as few nonzero entries as possible. FOCal Underdetermined System Solver (FOCUSS) is a classic example of this category, which uses a weighted minimum norm (MN) approach for sequentially reinforcing strong sources and suppressing the weak ones [15]. Other interesting algorithms are based on Iterative Reweighted Least-Squares (IRLS) methods, which are similar to FOCUSS and are based on iteratively computing weighted MN solutions with weights updated after each iteration [16, 17]. The homotopy method by Osborne [18] and LARS-LASSO algorithm [19, 20] (a variant of the homotopy method) are extremely powerful methods for solving the $l_1$ problem. Simple coordinate descent methods [21] or block wise coordinate descent, also called block coordinate relaxation [22], are also very successful strategies.

Following the discovery by Donoho and Candes et al. [23, 24] that sparsity could enable exact solution of ill-posed problems under certain conditions, there has been a tremendous growth of publications on efficient application of sparsity constraints for ill-posed problems [25-35]. An interesting study about the performance of several state of the art sparse approximation algorithms for solving EEG source localization problem has been done by, Zhang in [28]. In EEG, researches have used sparseness constraints in spatial, spatiotemporal and frequency domains to reflect the focal nature of the cortical activity. Ding et al. in [25, 26] have proposed an $l_1$-norm electromagnetic source imaging method by exploring sparseness in a transform domain. To efficiently incorporate the sparsity prior Chang et al. [27] have considered Laplacian and Spherical wavelet transform to solve the localization problem in MEG (magnetoencephalography). Wu et al. in [29] have proposed a matching pursuit based solution to the EEG inverse problem. While it produces better localization compared to the state-of-the art methods, the number of sources needs to be known a priori for the refinement of the localization in this method.

Despite the growing interest, the applicability of sparsity-enforcing priors for EEG source localization is still limited because of the significantly smaller number of electrodes compared to the typical number of virtual electric current dipoles in consideration. Importantly, it is well accepted in the literature that during any brain activity a group of virtual dipoles as opposed to a single dipole gets activated [36]. When such block structure is present in the signal, a trend in the field of sparse signal recovery is to exploit the structure of signal for better performance [1]. Block sparse signal recovery relies on the same mathematical paradigm of sparse signal recovery; however considers the signal as a collection of few nonzero blocks, where a block is a collection of consecutive nonzero entries. Relying on block structure, an number of algorithms exists in the literature that considers exploiting the correlation within the block for better performance -such as mixed $l_2/l_1$ program [37], block-OMP [38], block-CoSaMp [39], block sparse Bayesian learning (BSBL) [1]. Following the discovery of Zhang [40], in EEG, where the dictionary matrix is highly coherent, sparse Bayesian learning seems to outperform other state-of-the-art sparsity prior approaches. In this paper we evaluate the performance of the block sparse Bayesian learning algorithm proposed in [1] for EEG source localization in regard to realistic head model obtained from the segmentation of MRI images of the head. Brodmann areas [41] and automated anatomical labelling (AAL) [42] were used separately as blocks while generating the forward problem. Experiments were performed by varying the number of simultaneously active blocks and with and without the presence of noise.

**Recovery of Block Sparse Signal Using BSBL:**

The basic problem in sparse signal recovery is to recover a signal with only a few non-zero elements, from a small number of linear measurements [1]. Knowing most natural signals have rich structure, a trend in the field is to exploit structure of signals for better performance. A widely existing structure in natural signals is block structure [38]. Its mathematical model is given by:

$$\boldsymbol{\Phi} = \boldsymbol{K}\boldsymbol{J} + \boldsymbol{n}, \tag{1}$$

where the signal $\boldsymbol{J}$ has a block structure:

$$\boldsymbol{J} = [\underbrace{J_1, \dots, J_{d_1}}_{\boldsymbol{J}^{1^T}}, \dots, \underbrace{J_{d_{g-1}+1}, \dots, J_{d_g}}_{\boldsymbol{J}^{g^T}}]^T, \tag{2}$$

$\boldsymbol{\Phi} \in \mathbb{R}^{N_E \times 1}$ is the measurement vector consisting of $N_E$ measurements, $\boldsymbol{K} \in \mathbb{R}^{N_E \times M}$ is a known matrix ($N_E << M$) and $\boldsymbol{n}$ is the noise vector. We want to recover $\boldsymbol{J} \in \mathbb{R}^{M \times 1}$.. The block partition in (2), $d_1, \dots, d_g$ are not necessarily identical and among the $g$ blocks only a few blocks are nonzero. It is shown in [1, 43, 44], if such block partition can be exploited, then under certain conditions, the number of measurements required to recover $\boldsymbol{J}$ can be further reduced.

Relying on the internal block structure of the signal, efficient recovery algorithms are proposed in [1] by learning and exploiting the intra-block correlation of the signal. Out of the two algorithms proposed in [1] we consider the algorithm that is directly derived from the bSBL framework [45] and requires a priori knowledge of the block partition.

The BSBL [1] assumes each block $\boldsymbol{J}^i \in \mathbb{R}^{d_i \times 1}$ satisfies a parameterized multivariate Gaussian distribution:

$$p(\boldsymbol{J}^i) \sim \mathcal{N}(0, \gamma_i \boldsymbol{B}_i); i = 1, \dots, g,$$

where $\gamma_i$ is a non-negative parameter, $\boldsymbol{B}_i \in \mathbb{R}^{d_i \times d_i}$ is a positive definite matrix, capturing the correlation structure of the $i$-th block. Further by assuming the blocks are mutually uncorrelated, the prior for $\boldsymbol{J}$ is given by $p(\boldsymbol{J}) \sim \mathcal{N}(\boldsymbol{0}, \boldsymbol{\Psi}_0)$, where $\boldsymbol{\Psi}_0$ is a block diagonal matrix with each principal block given by $\gamma_i B_i$. The noise vector is assumed to satisfy

$p(\boldsymbol{n}) \sim \mathcal{N}(\boldsymbol{0}, \lambda I)$, where $\lambda$ is a non-negative scalar. The posterior of $\boldsymbol{J}$ is given by $p(\boldsymbol{J}|\boldsymbol{\Phi}) = \mathcal{N}(\boldsymbol{\mu}_J, \boldsymbol{\Psi}_J)$ with $\boldsymbol{\mu}_J = \boldsymbol{\Psi}_0 K^T (\lambda I + K \boldsymbol{\Psi}_0 K^T)$ and $\boldsymbol{\Psi}_J = (\boldsymbol{\Psi}_0^{-1} + \frac{1}{\lambda} K^T K)^{-1}$. Once the hyperparameters $\lambda, \gamma_i, \boldsymbol{B}_i$ (for $i = 1, \dots, g$) are estimated the Maximum-A-Posterior (MAP) estimate of $\boldsymbol{J}$ can be directly obtained from the mean of the posterior. Following the Expectation Maximization (EM) method in [45], the iterative algorithm of the BSBL methodology can be described as follows:

$$\boldsymbol{\mu}_J \leftarrow \boldsymbol{\Psi}_0 K^T (\lambda I + K \boldsymbol{\Psi}_0 K^T)^{-1} \boldsymbol{\Phi}$$

$$\boldsymbol{\Psi}_J \leftarrow \boldsymbol{\Psi}_0 - \boldsymbol{\Psi}_0 K^T (\lambda I + K \boldsymbol{\Psi}_0 K^T)^{-1} K \boldsymbol{\Psi}_0$$

$$\lambda \leftarrow \frac{||\boldsymbol{\Phi} - K \boldsymbol{\mu}_J||_2^2 + \lambda[M - \mathrm{Tr}(\boldsymbol{\Psi}_J \boldsymbol{\Psi}_0^{-1})]}{N_E}$$

$$\gamma_i \leftarrow \frac{1}{d_i} \mathrm{Tr}[\boldsymbol{B}_i^{-1}(\boldsymbol{\Psi}_J^i + \boldsymbol{\mu}_J^i (\boldsymbol{\mu}_J^i)^T)]; \; \forall_i \in 1:g$$

$$\boldsymbol{B}_i \leftarrow \frac{\boldsymbol{\Psi}_J^i + \boldsymbol{\mu}_J^i (\boldsymbol{\mu}_J^i)^T}{\gamma_i}; \; \forall_i \in 1:g$$

where $\boldsymbol{\mu}_J^i$ corresponds to the $i$-th block in $\boldsymbol{\mu}_J$ and $\boldsymbol{\Psi}_J^i$ corresponds to the $i$-th principal diagonal block in $\boldsymbol{\Psi}_J$. Once the algorithm converges, the estimate of $\boldsymbol{J}$ is given by $\boldsymbol{\mu}_J$.

**Data Model and Assumptions:**

A realistic head model was obtained from segmentation of MRI images of the head and includes four major compartments, namely scalp, skull, cerebrospinal fluid (CSF) and brain, with the following relative conductivity values [36]: $\sigma_{\text{scalp}}=1$, $\sigma_{\text{skull}}=0.05$, $\sigma_{\text{CSF}}=5$, $\sigma_{\text{brain}}=1$. The source space was constructed by dividing the head model into cubes with a size of $15 \times 15 \times 15$ mm$^3$ and considering possible current dipoles only at the center of those cubes that consisted of at least 60% of gray matter. This segmentation procedure resulted in 230 dipole positions[1]. To be consistent with the BSBL framework these 230 dipoles were clustered into blocks based on Brodmann areas [41] and automated anatomical labelling (AAL) [42] separately.

---

[1] 230 dipole positions result 690 unknowns in total

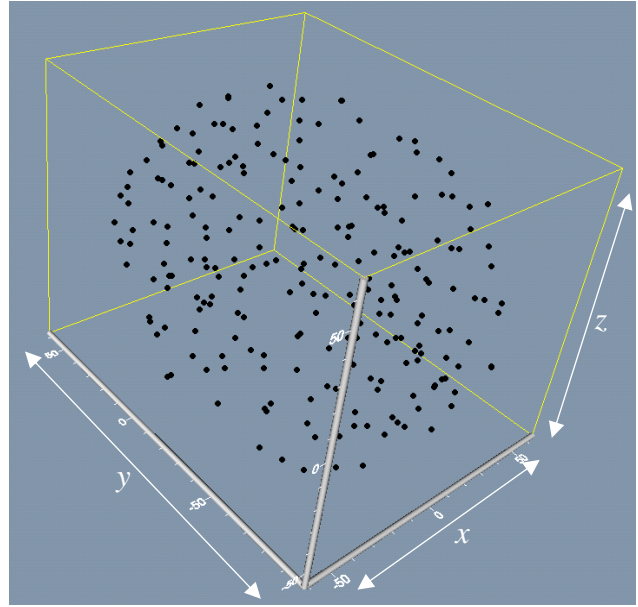

Figure 1: Positions of the 230 dipoles.

Brodmann's map of the human cortex contains 86 cytoarchitectonic areas in total (considering the left and right half as separate areas), however for the considered head model we were able to form only 67 clusters. We suspect this is due to the coarse sampling of the head model, which is typical in EEG; also due to the percentage of gray matter associated with a dipole; which resulted in zero dipoles in small Brodmann areas. For the AAL based clustering we used the template of the MRIcro [46] package. The template consists of 116 areas in the standard MNI space. However for the considered head model we were able to form only 97 clusters. The resultant decrease of the number of clusters is due to the coarse sampling of the head model in EEG compared to fMRI and also due to the percentage of gray matter associated with a dipole; which resulted in zero dipoles in small anatomical areas.

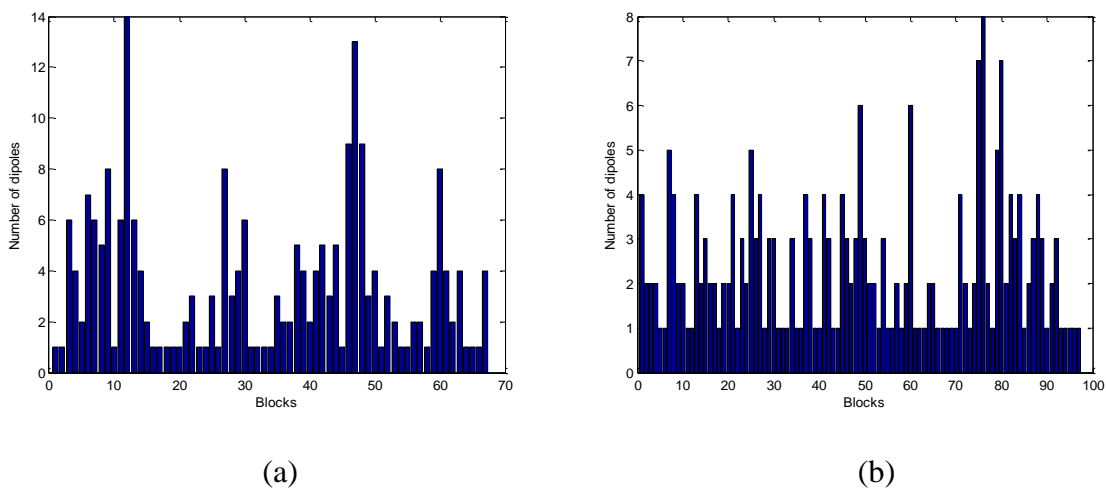

(a)                                              (b)

Figure 2: Number of dipoles per block - (a) Brodmann area based clustering, (b) AAL based clustering.

**Source Localization:**

To localize the sources of electrical activity or more specifically to localize the blocks that are active, we solve the inverse problem based on BSBL [1] methodology. The solution $\hat{J} \in \mathbb{R}^{690 \times 1}$ is a component vector representing current sources at 230 locations within the brain volume with three directional (i.e. $x$, $y$, $z$ directions) components per location. The $x$, $y$ and $z$ components are used to calculate the magnitude, $d = \sqrt{x^2 + y^2 + z^2}$ of the current density for each of the 230 dipoles. The magnitude of each of the blocks (i.e. 67 blocks for Brodmann area based clustering or 97 blocks for AAL based clustering) is computed by taking the summation of the magnitudes of all the dipoles that belongs to the block. From the magnitudes of the blocks $|J_{Block}| \in \mathbb{R}^{N_B^2 \times 1}$, the maximum magnitude, $|J_{Block}|_{\max}$ is determined, any block with a magnitude larger than or equal to $t|J_{Block}|_{\max}$, is considered to be active. The threshold, $t$ is experimentally set to 1/3 of $|J_{Block}|_{\max}$.

**Experiments:**

Experiments were conducted by varying the number of simultaneously active blocks for the EEG headset configuration shown below. Along with considering 2 different block structures based on Brodmann areas and AAL, we considered two different cases. In case-I we considered all the dipoles of the activated block follow the same orientation and in case-II we considered the dipoles of the activated block are randomly oriented. In both of the cases we considered all the dipoles of the block have same magnitudes irrespective of their orientations.

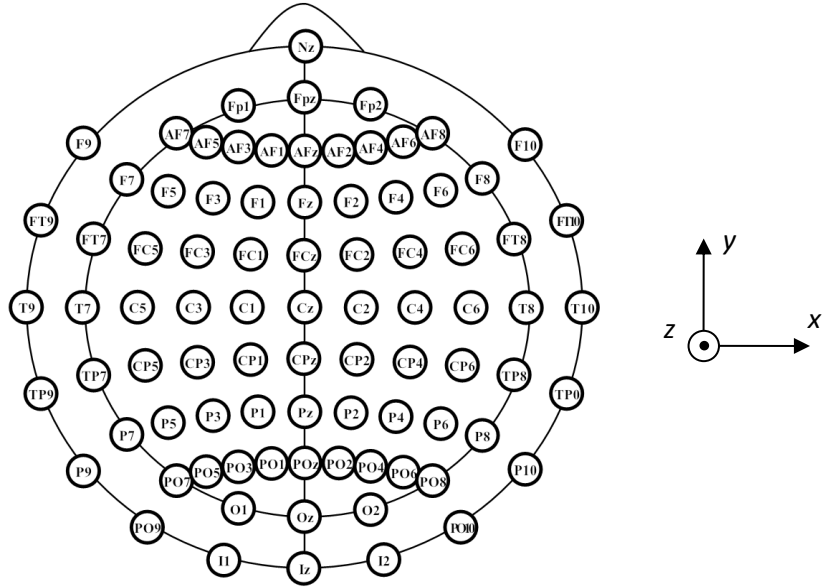

Figure 3: Schematic representation of the electrodes positions in the 71-electrodes setup.

---

² $N_B = 67$ for BA based clustering or 97 for AAL based clustering

Error distance (ED) [Yao05] was used to analyse the reconstruction performance.

$$\text{ED} = \frac{1}{N_I} \times \sum_{i \in I}^{N_I} \min_l ||r_i - s_l|| + \frac{1}{N_L} \times \sum_{l \in L}^{N_L} \min_i ||s_l - r_i|| \tag{3}$$

Here $s_l$ and $r_i$ are the centroid of the actual and estimated active blocks respectively. $N_I$ and $N_L$ are the total numbers of estimated and the undetected sources (i.e. active blocks) respectively. The first term of equation (3) calculates the mean of the distance from each estimated source to its closest real source, and the corresponding real source is then marked as detected. All the undetected real sources made up the elements of the data set $L$ and thus the second term of the equation calculates the mean of the distance from each of the undetected sources to the closest estimated source.

## A) Noise-Free Simulation:

### i) Single Block Activation:

In this case to generate the forward problem, a single block was made active (more specifically all the dipoles in the block were active) and the corresponding potentials on the electrodes were calculated. The experiment was conducted for all the blocks activated sequentially one at a time. For the experiment we call a 'success' if the activated block is exactly localized in the reconstructed signal. The success rate is computed as $\frac{\sum_{i=1}^{N_{Blocks}}(ED^i==0)}{N_{Blocks}}$. For the unsuccessful cases we computed the 'error distance' between the actual and the reconstructed signal. Taking into account the unsuccessful cases, we computed the mean error distance as $\frac{\sum_{i=1}^{N_{Blocks}}(ED^i)}{N_{unsuccessful}}$ and the standard deviation of the mean error distance. While generating the forward problem the dipole orientations were generated randomly, therefore, the whole experiment was repeated 1000 times. Figures 4 and 5 represent the average of 1000 such findings.

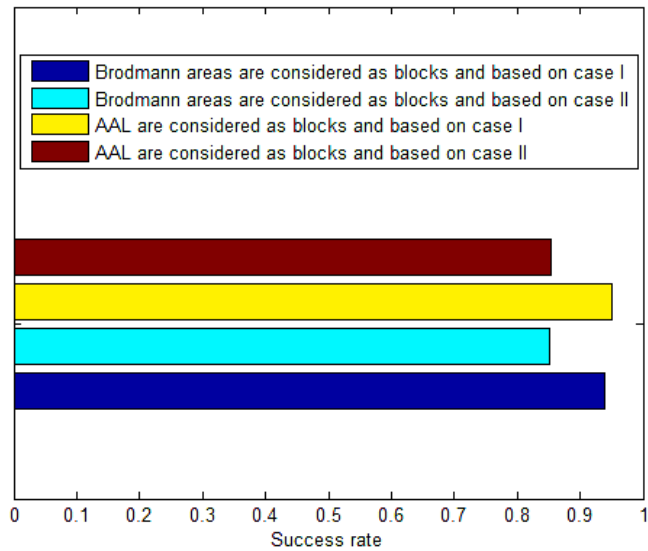

Figure 4: Success rate of BSBL method [1] for a single active block. Blocks are defined based on Brodmann areas and AAL separately.

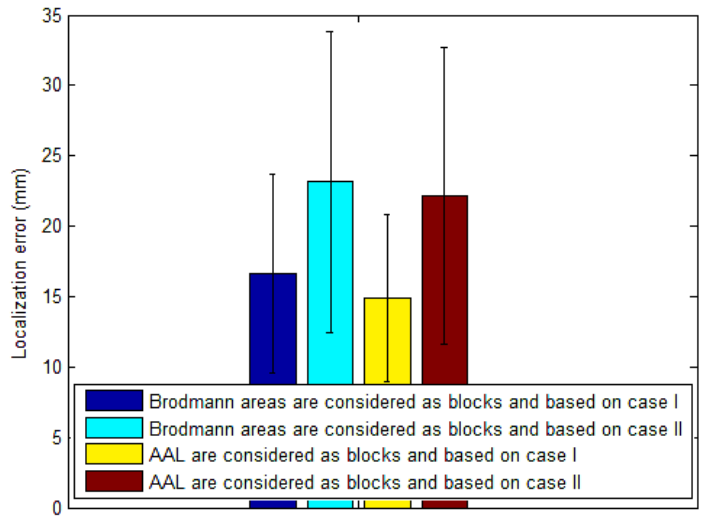

Figure 5: Localization error for the unsuccessful cases – bars showing the mean error distance and whiskers showing the standard deviation of the mean error distance.

From the results it is observable that when a single block of dipoles is active, it is very likely to be exactly localized by the BSBL method. We also observe that when all the dipoles of the active block follows the same orientation (i.e. case-I) the chances of exact localization are higher in comparison to when all the dipoles of the active block follows the different orientation (i.e. case-II). Interestingly, the performance of exact localization is marginally better for AAL based clustering than for Brodmann area based clustering, although Brodmann area based clustering produces a system, which is comparatively less underdetermined. In our understating, clustering based on Brodmann area produces few cluster with significantly large number of dipoles in comparison to others and when all the dipoles of such a large cluster is active, in addition with the actual active block few additional blocks are detected in the reconstructed signal. These unwanted blocks result the slightly decreased performance for Brodmann area based clustering in comparison to AAL based clustering.

### ii) Multi Block Activation:

In this case $S$ ($S > 1$) blocks were activated simultaneously. From the total $C_S^{N_B} = \frac{N_B!}{S!(N_B - S)!}$ possible combinations of blocks, one combination was chosen randomly. We determined the block that had the minimum number of dipoles and considered only that many active dipoles for all the blocks. This was to make sure all the blocks that were made simultaneously active have same magnitude. When we were to activate a fraction of the dipoles within a block rather than all the dipoles, one dipole of the block was chosen randomly and the rest of the active dipoles (based on the minimum number of dipoles) were adjacent[3] to the chosen dipole. We computed the error distance between the actual and reconstructed signal and claimed a "success" if the computed error distance was zero. We did the experiment 1000 times (varying the combinations of activated blocks and the orientation of the average dipole moment) and the success rate was computed over these 1000 runs as $\frac{\sum_{i=1}^{1000}(ED^i == 0)}{1000}$. For the unsuccessful cases we computed the mean error distance, $\overline{ED} = \frac{\sum_{i=1}^{1000} ED^i}{N_{unsuccessful}}$ and the standard deviation of the mean error distance. We followed this procedure for each value of *S,* and for both Brodmann areas and AAL. Figures 6 and 7 show the average of 1000 such findings.

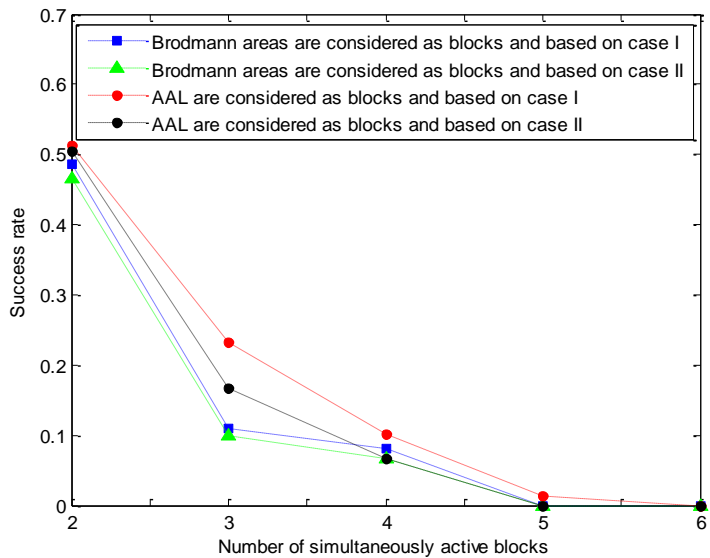

Figure 6: Success rate of BSBL method [1] for simultaneously active multiple blocks.

---

[3] Closely located

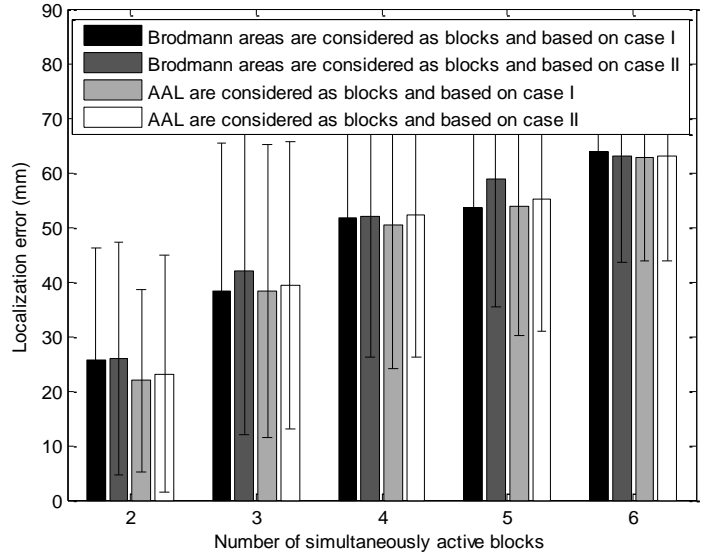

Figure 7: Localization error for the unsuccessful cases in the presence of simultaneously active multiple blocks. Bars showing the mean error distance and whiskers showing the standard deviation of the mean error distance.

From the results we observe that localization accuracy decreases with the increased number of simultaneously active blocks. As expected, better localization accuracy is obtained for case-I compared to case-II. Interestingly, AAL based clustering performs compared to Brodmann area based clustering. To our understanding, the variance between the cluster sizes is significantly higher in Brodmann area based clustering compared to AAL based clusters (see Figure 2), which makes reconstruction more challenging for  Brodmann are based clustering.

### B) Experiment with Noisy Data:

All earlier experiments were performed in absence of noise. However, experimental data are likely be contaminated with "noise" from various sources, including measurement noise and background brain activity [47]. We therefore investigated the effect of simulated pseudo-random measurement noise superimposed on the measured scalp potentials.

### i) Single Block Activation:

First, a single block was considered active and the corresponding potentials on the electrodes were calculated. We then added variable amounts of noise to each of the electrode potentials. The noise vector $\boldsymbol{n}$ was generated based on the SNR (signal-to-noise ratio) defined below:

$$\text{SNR (in dB)} = 20 \log_{10} \frac{||\boldsymbol{KJ}||_2}{<||\boldsymbol{n}||_2>} \qquad (4)$$

where <...> designates the mean value over a statistical ensemble.

In this study we increased the SNR from 5 dB to 30 dB with 5dB increments. The experiment was conducted for all the blocks activated sequentially one at a time. For the experiment we call a 'success' if the activated block is exactly localized in the reconstructed signal. The success rate is computed. We computed the success rate as $\frac{\sum_{i=1}^{N_{Blocks}}(ED^i==0)}{N_{Blocks}}$ and the mean error distance (considering only unsuccessful cases). Since we were adding a random amount of noise, we did 1000 repetitions of the experiment. For each run we computed success rate and mean error distance and standard deviation of the mean error distance. Figures 8-9 show the average of 1000 such findings.

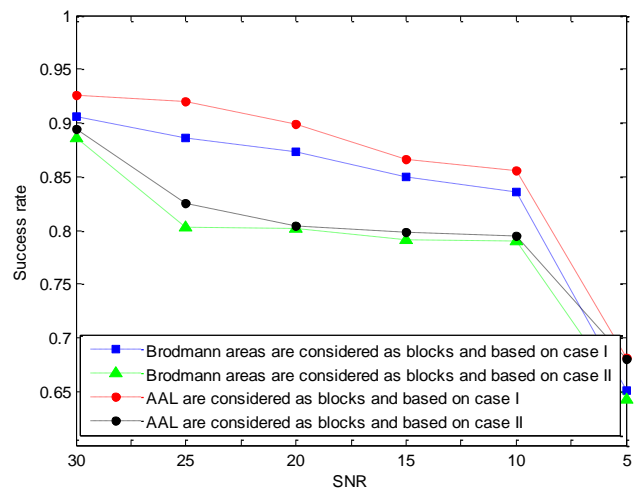

Figure 8: Success rate of BSBL method [1] for different levels of noise considering a single active block at a time.

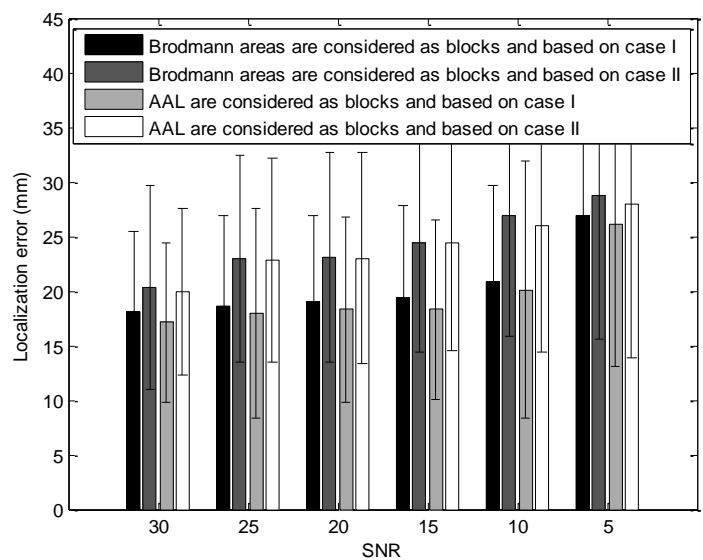

Figure 9: Localization errors (considering only unsuccessful cases) with respect to different noise levels.

From the results we observe that the localization accuracy decreases with the increased amount of noise. We also observe that better localization accuracy is perceivable for case-I in comparison to case-II. Finally, similar to the previous experiments, performance of localization is better for AAL based clustering compared to Brodmann clustering.

### ii) Multiple Blocks Activation:

In this case S (S>1) blocks were activated simultaneously. From the total of $C_S^{N_B} = \frac{N_B!}{S!(N_B-S)!}$ possible combinations of blocks one combination was chosen randomly. We determined the block that had the minimum number of dipoles and considered only that many active dipoles for all the blocks as we generated the forward problem. This would ensure that all the blocks that were made simultaneously active have same magnitude. When we were to activate a fraction of the dipoles within a block rather than all the dipoles, one dipole of the considered block was chosen randomly and the rest of the active dipoles were adjacent to the chosen dipole. We computed the error distance between the actual and reconstructed signal and claimed a "success" if the computed error distance was zero. We performed the experiment 1000 times by varying the combinations of activated blocks and the orientation of the average dipole moment. We computed the success rate as $\frac{\sum_{i=1}^{1000}(ED^i==0)}{1000}$, for the unsuccessful cases we computed the mean error distance and standard deviation of the mean error distance. Since we were adding random amount of noise, we did 1000 repetitions of the experiment. Figures 10-11 represent the average of 1000 such findings.

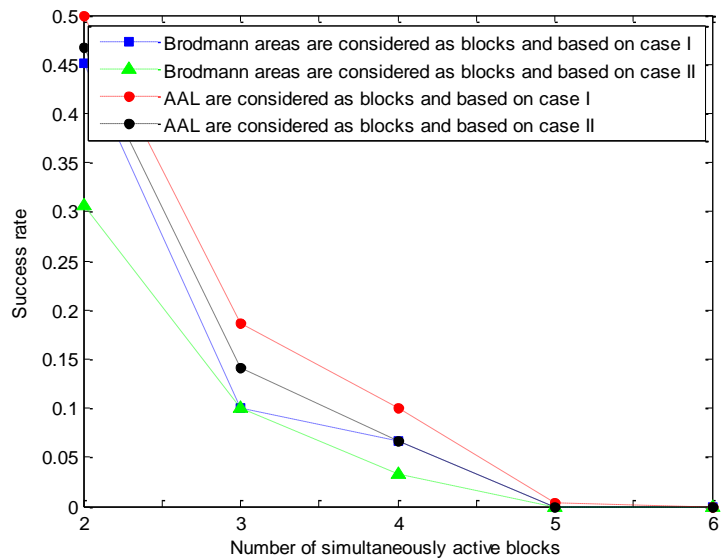

Figure 10: Success rate of BSBL method [1] for simultaneously active multiple blocks in the presence of 20 dB noise.

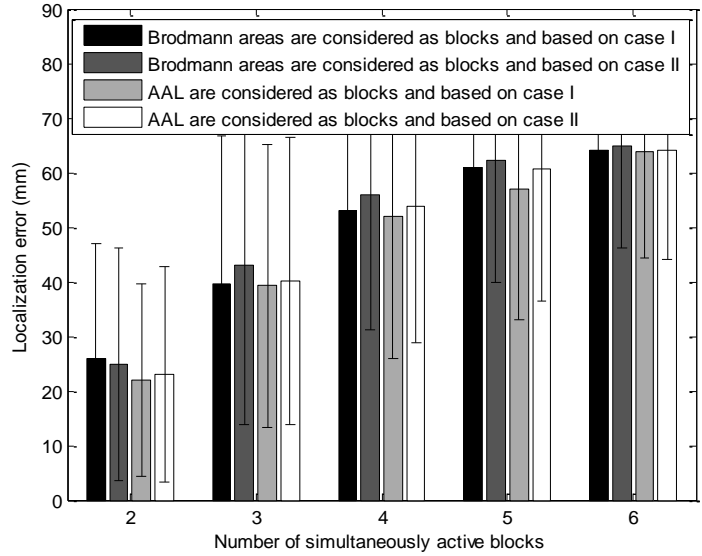

Figure 11: Localization error for the unsuccessful cases in the presence of simultaneously active multiple blocks. Experiments were performed in the presence of 20 dB noise.

From the results we observe that localization accuracy decreases with the increased number of simultaneously active blocks. Similar to the noiseless scenario, here also the performance of localization is better for AAL based clustering than Brodmann area based clustering, with a fall in localization accuracy for both clustering.

**Discussion and Conclusions:**

In this study we have quantitatively analyzed and evaluated the performance of BSBL [1] method for EEG source localization on a realistic head mode, with and without the presence of noise. Exploiting the internal block structure, BSBL method solves severely underdetermined problem more efficiently compared to the methods that do not consider block structure [1]. The consideration of block structure is a natural choice for EEG, since it is already widely accepted by the community that a group of dipoles rather than a single dipole gets activated for certain brain activity. BSBL [1] can return a block sparse signal without and with the availability of a priori knowledge of the block partition. Obviously with the availability of a priori knowledge of the block partition, the recovery performance is better. Thus, for a successful application of the BSBL method we group the dipoles inside human brain and then consider each group as a block. Since the full functionality of different parts of the human brain is still not known completely, we group dipoles based on Broadmann areas and AAL, which is a cytoarchitectural classification method and an anatomical classification method, respectively. Numerous recent studies have unveiled the potential suitability of Broadmann segmentation of the brain as a tentative basis for a functional classification relevant to EEG [48]. Similarly, AAL map [42] is frequently used in fMRI to describe the region of interest [49], which indicates its potential to become a tentative basis for functional classification relevant to EEG.

The BSBL code implemented in MATLAB was downloaded from the author's website. Out of the two algorithms, we considered the one, which requires a priori knowledge of the block partition. We evaluated the performance of the BSBL method by varying the number of simultaneously activated blocks and in regard to Brodmann areas and AAL.

The results show when there is only one active block with all dipoles inside the block follow the same orientation, BSBL method produces exact localization for around 95% cases in absence of noise. With the presence of noise or when the dipoles within the active block are randomly oriented, the localization accuracy decreases. With the presence of multiple active blocks the percentage of exact localization decreases and drops down to 0 for 5 simultaneously active blocks. AAL based clustering/partitioning produces better results in comparison to Brodmann area based clustering, because of their comparative uniform cluster sizes. For Brodmann area based clustering, when all the dipoles within a block follow the same orientation, the overall localization error[4] is about 14 mm for 2 simultaneously active blocks. For 3 and 4 simultaneously active blocks the overall localization errors are 35 mm and 53 mm respectively. For AAL based clustering the overall localization error is about 11 mm for 2 simultaneously active blocks. For 3 and 4 simultaneously active blocks the overall localization errors are 31 mm and 47 mm respectively.

Although BSBL is a popular strategy to solve the severely underdetermined problem in EEG source localization, it is still insufficient to produce a reliable source localization. It is rather suitable (produces <15mm error) when number is simultaneously active blocks is very small (maximum 2); When the number of simultaneous active blocks increases (greater than 2), the overall localization error drastically increases (> 30 mm for 3 blocks). However, simultaneously active blocks with unequal strength present a highly complex scenario and poorer source localization performance is not surprising. Again, with the presence of noise, and other realistic factors, the localization performance is logically subject to degrade.

---

[4] (1−Success Rate)×Localization Error